\documentclass[twocolumn]{aastex631}

\newenvironment{Contfigure}{%
\renewcommand{\thefigure}{\arabic{figure}}%
\addtocounter{figure}{-1}%
\begin{figure}}{%
\renewcommand{\thefigure}{\arabic{figure}}%
\end{figure}}

\graphicspath{{./}{figures/}}

\shorttitle{Ancient nova shells of RX~Pup indicate evolution of mass transfer rate}
\shortauthors{I\l{}kiewicz et al.}

\begin{document}

\title{Ancient nova shells of RX~Pup indicate evolution of mass transfer rate}

\author[0000-0002-4005-5095]{Krystian I\l{}kiewicz}
\affiliation{Astronomical Observatory, University of Warsaw, Al. Ujazdowskie 4, 00-478 Warszawa, Poland}
\affiliation{Centre for Extragalactic Astronomy, Department of Physics, Durham University, DH1 3LE, UK}

\author[0000-0003-3457-0020]{Joanna Miko\l{}ajewska}
\affiliation{Nicolaus Copernicus Astronomical Center, Polish Academy of Sciences, Bartycka 18, 00716 Warsaw, Poland}

\author[0000-0003-0155-2539]{Michael M. Shara}
\affiliation{Department of Astrophysics, American Museum of Natural History, CPW \& 79th street, New York, NY 10024-5192, USA}

\author[0000-0001-6251-0573]{Jacqueline K. Faherty}
\affiliation{Department of Astrophysics, American Museum of Natural History, CPW \& 79th street, New York, NY 10024-5192, USA}

\author[0000-0001-5387-7189]{Simone Scaringi}
\affiliation{Centre for Extragalactic Astronomy, Department of Physics, Durham University, DH1 3LE, UK}

\affiliation{INAF-Osservatorio Astronomico di Capodimonte, Salita Moiariello 16, I-80131 Naples, Italy}


\begin{abstract}
RX~Pup is a symbiotic binary which experienced a nova outburst in the 1970's. Here we report a discovery of a $\sim$1300 year old nova shell around the system and a possible detection of a $\sim$7000~year old nova shell. Together with the nova shell ejected in the 1970's this makes RX~Pup the first system with three nova shells observed. This triad of eruptions suggests a change in the nova recurrence time. The most likely explanation is an alteration in the mass transfer rate attributed to evolutionary changes of the mass-donor in the system. Notably, comparative analyses with theoretical models indicate an increase in the average mass transfer rate by a factor of three over the past 10,000 years. This makes RX Pup a unique system, which allows us to probe millenium-scale evolution of mass transfer rates in binary systems.

\end{abstract}



\section{Introduction} \label{sec:intro}

Symbiotic stars (SySt) are interacting binary system in which a white dwarf (WD) or a neutron star is accreting matter from a red giant companion. They can experience classical nova (CN) eruptions, which occur on the surface of the WD when enough material is accumulated and a thermonuclear runaway is triggered \citep{1974ApJS...28..247S,1978A&A....62..339P}. The material ejected during the eruption forms a nova shell that can be observed for  millenia after the outburst \citep{2017Natur.548..558S,2020A&A...641A.122T,2024MNRAS.529..212S}. 

CNe in SySt are either extremely slow novae (usually called symbiotic novae; SyNe)  with outbursts going on for decades, or very fast recurrent novae (SyRNe) with very short, $\sim$several days, timescales and recurrence time of $\sim$several years or decades. The differences in outburst behavior seem to reflect different WD masses: very high in SyRNe, close to the Chandrasekhar limit, and much lower in SyNe \citep{2010arXiv1011.5657M}.

RX~Pup is SySt with an orbital period of at least 200~years and a extreme Mira donor pulsating with a period of $\sim$578~days \citep{1999MNRAS.305..190M}. RX~Pup experienced a CN outburst that started in the 1970’s. The outburst evolution in the HR diagram was consistent with a slow SyN on a $\sim$0.8~M$_\odot$ WD \citep{1999MNRAS.305..190M}. RX Pup  produced a nova shell that was visible even during outburst \citep{1989ApJ...337..514H,1999MNRAS.305..190M,1990ApJ...357..231P,2000A&A...363..671C,2002AIPC..637...42M}.

\citet{1999MNRAS.305..190M} suggested that another nova outburst occurred in the 1890’s, however it could have been as well a high state due to enhanced accretion rate,  similar to that observed in the late 1990’s \citep{2002AIPC..637...42M}. We also note that recurrent CN outbursts have been recently reported for two other SySt, LMC~S154 \citep{2019A&A...624A.133I} and V618~Sgr \citep{2023MNRAS.523..163M}, both showing relatively slow outbursts more similar to SyNe than to SyRNe, suggesting less massive WDs. A possible solution to this conundrum maybe their much higher mass accretion rates than those of CNe with unevolved donors which would significantly shorten the CN recurrence time. In addition, evolved donors are bright in the optical red and near infrared range which makes relatively easy to measure radial velocities from absorption lines, and to derive reliable masses of the binary components.
For this reason SyNe and SyRNe are a valuable tool for testing theoretical models of CN outbursts.

In this study, we report the discovery of an additional nova shell associated with RX Pup originating from a prior outburst, as well as a possible, even older nova shell.

\section{Observations} \label{sec:obs}
\subsection{Narrow-band imaging}
\begin{figure*}
\includegraphics[width=1.0\hsize]{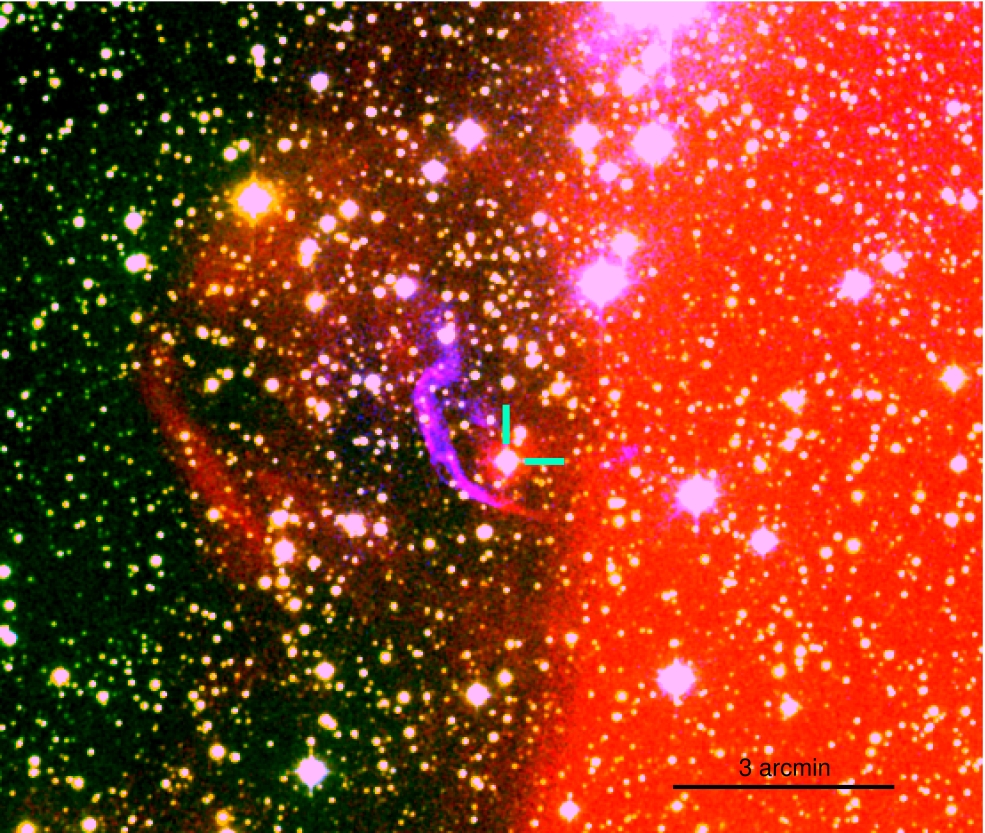}
\caption{False-colour image of RX~Pup  with [O~III] represented in blue, $R$ in green and H$\alpha$ in red colour. North is up, east is left. RX Pup is in the center of the image, indicated with turquise ticks. The entirety of the image west of RX~Pup is dominated by H$\alpha$ emission of a unrelated nearby H~II region. The $\sim$1300 year old nova shell is seen as a bright [O~III] and H$\alpha$ emission 43 arcsec east of RX Pup. A possible $\sim$7000 year old shell is seen as a H$\alpha$ bright shell centered on RX~Pup and with a radius of 240 arcsec.}
\label{fig:colored}
\end{figure*}

We observed RX~Pup with the 1-metre Swope telescope at Las Campanas, Chile on 06-07 December 2016. The telescope was equipped with a 4K$\times$4K E2V CCD231-84 camera, resulting in a 30$\times$30 arcmin field of view. The observations were carried out in [O~III], H$\alpha$, and $R$ filters. This included 2$\times$1200s and 2$\times$2400s frames in H$\alpha$, 1$\times$1200s and 3$\times$2400s in the [O~III] filter, and 2$\times$45s and 3$\times$60s in the $R$ filter.  Each exposure was randomly offset to average out differences between the four independent detector amplifiers. The observations were reduced using standard $IRAF$ procedures. The astrometry was applied using the \textit{Astrometry.net} \citep{2010AJ....139.1782L}. The observations were combined using the \textit{SWarp} code \citep{2002ASPC..281..228B}.  A false-colour image of RX~Pup is presented in Fig.~\ref{fig:colored}.

\begin{figure*}[h!]
\includegraphics[width=0.49\hsize]{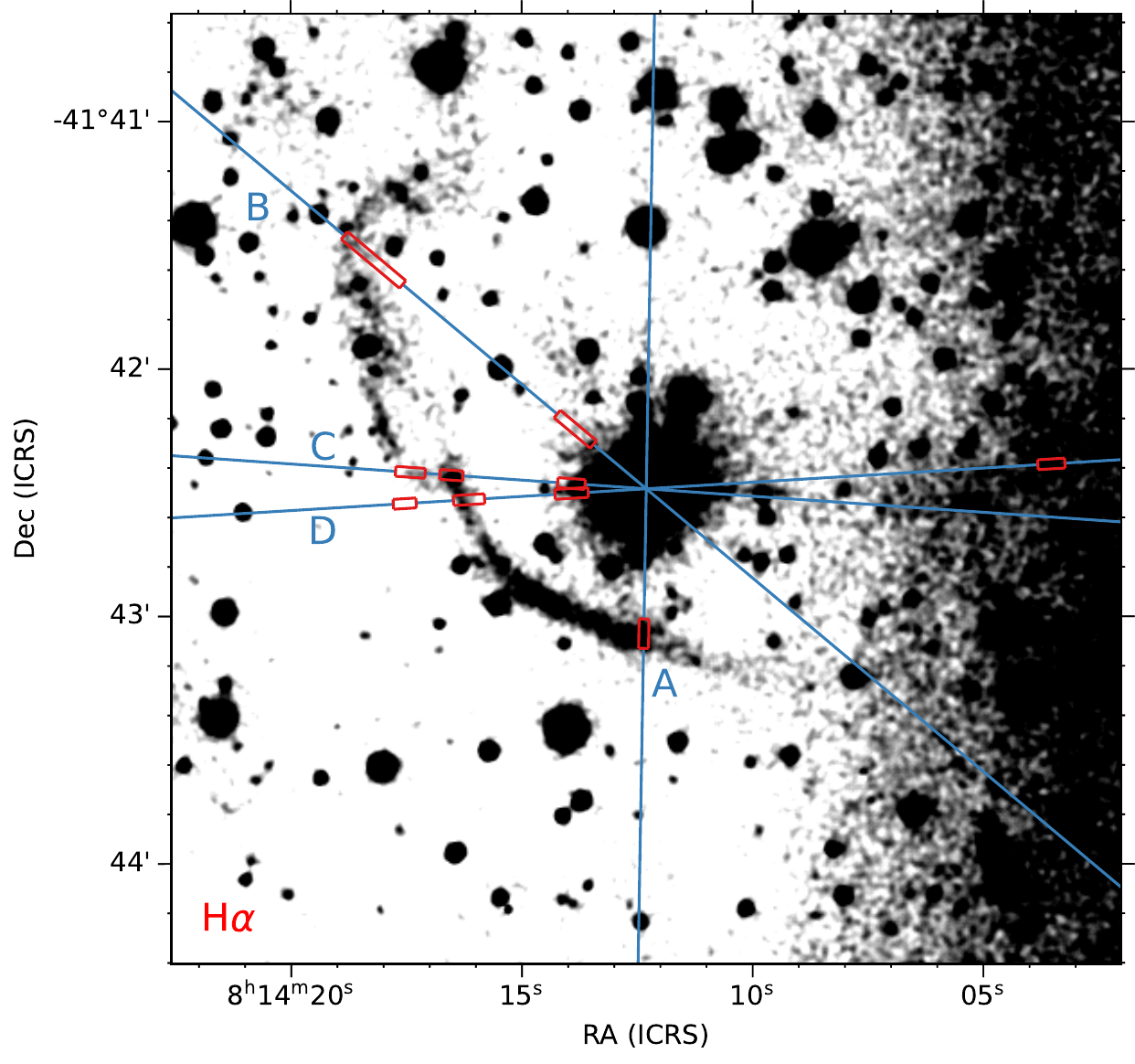}
\includegraphics[width=0.49\hsize]{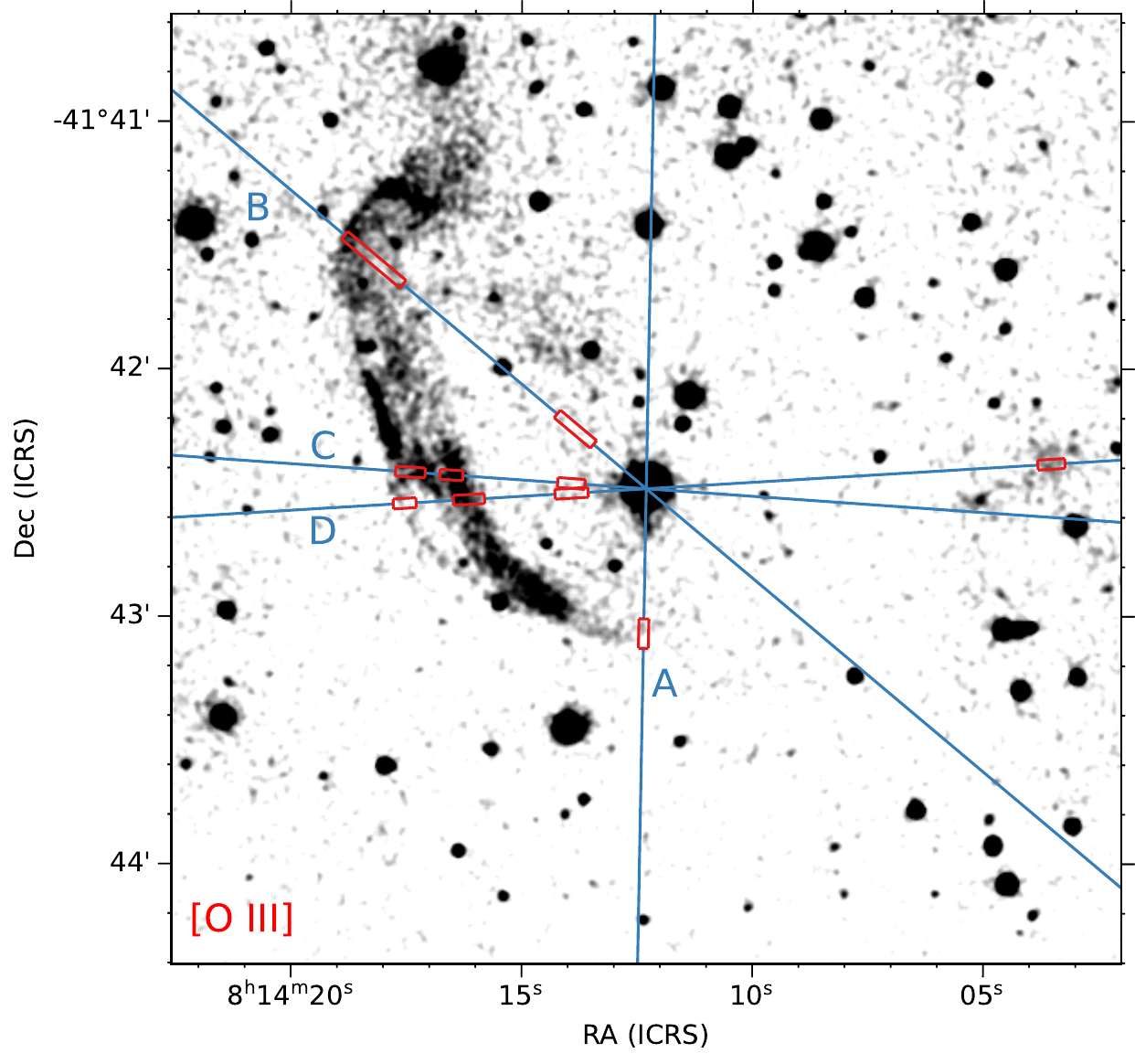}
\includegraphics[width=0.49\hsize]{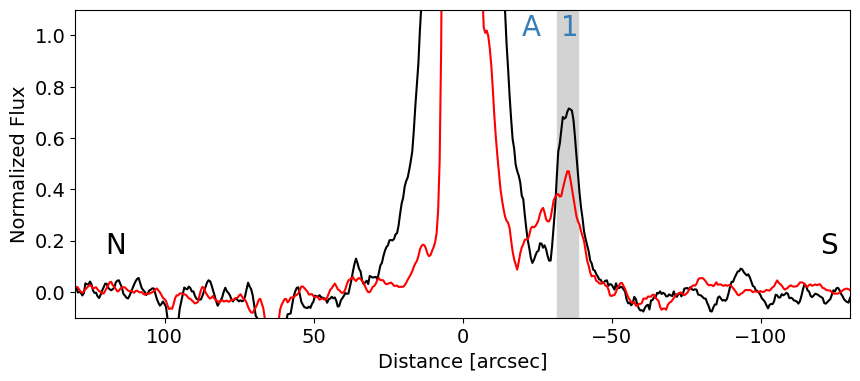}
\includegraphics[width=0.49\hsize]{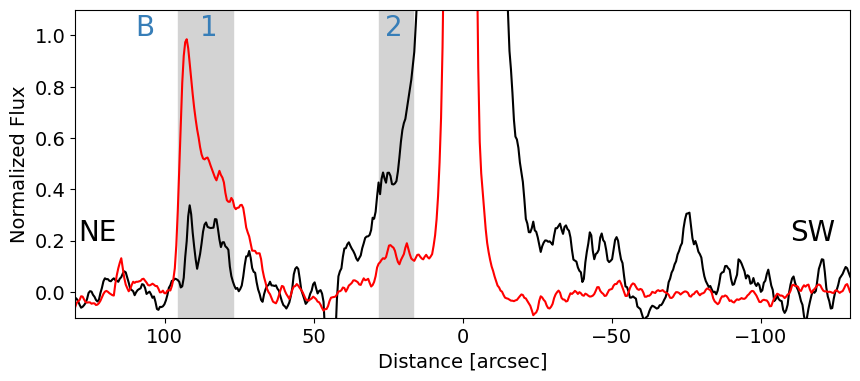}
\includegraphics[width=0.49\hsize]{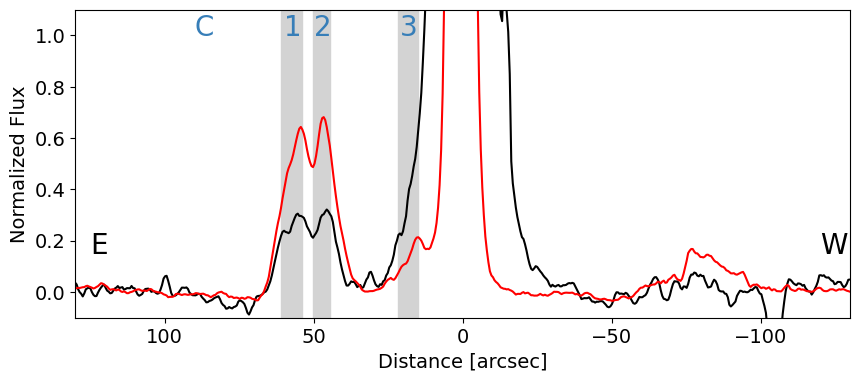}
\includegraphics[width=0.49\hsize]{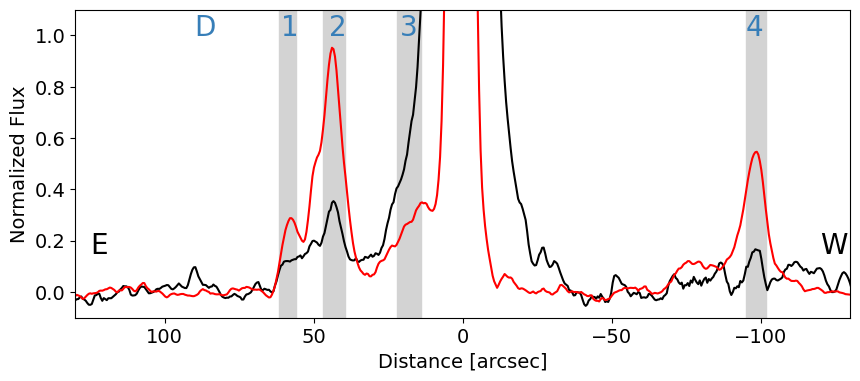}
\caption{  Top two panels:  H$\alpha$ and [O~III] images of extended emission regions around RX~Pup. Slit trails are marked with blue lines and named A--D with positional angles of 179, 50, 86 and 93.5~degrees respectively. All slits were centered at the position of RX~Pup. The identified extended emission regions along slit positions are marked with red rectangles. Bottom four panels: Observed fluxes along slit positions (A--D) as a function of distance from RX~Pup that were smoothed with a five pixel wide tophat filter. The direction on the sky from RX~Pup is marked with black letters. The observed fluxes were at H$\alpha$ (black line) and [O~III]~5007 (red line) wavelengths. For each slit position the identified extended emission regions are marked with a gray area. The extended emission regions are then numbered. The final assigned slit position name and the extended region number are marked with blue symbols. }
\label{fig:slits}
\end{figure*}

\subsection{Spectroscopy}

Spectroscopy was carried out with the Southern African Large Telescope (SALT; \citealt{2006SPIE.6269E..0AB,2006MNRAS.372..151O}) equipped with the Robert Stobie Spectrograph \citep{2003SPIE.4841.1463B,2003SPIE.4841.1634K} and a PG0900 grating yielding a resolution of 6.1\AA. The background subtraction was performed by fitting a low order polynomial to the sky at each individual wavelength. This subtracted both the telluric emission as well as emission from the nearby H~II region, leaving only the star spectra and more localized extended emission. Moreover, a low order polynomial was fitted to spectra at each position on the slit in order to minimize noise from background stars and scattered light from RX~Pup itself. The observations were carried out at four positional angles and were centered at RX~Pup (Fig.~\ref{fig:slits}). Each positional angle was chosen so that an additional star was on the slit. The distance on the sky between the additional star and RX~Pup was used to calibrate the distance of features observed along the slit. Spectrophotometric calibration was performed using a standard star. However, absolute calibration was impossible due to the variable aperture of SALT.  The list of spectroscopic observations is presented in the Appendix. 

In order to identify features of the extended nebula around RX~Pup, we extracted fluxes at H$\alpha$ and [O~III]~5007 wavelengths along the slit. For each slit position we identified extended emission regions using the fluxes along the slit and compared them to imaging. The emission regions are marked with red rectangles in top two panels of Fig.~\ref{fig:slits} and were named and numbered as shown in the bottom four panels. For each of these regions a median spectrum was extracted. The spectra are presented in the Appendix.

The measured emission line fluxes for each region are presented in Tab.~\ref{tab:fluxes}. The presented emission line fluxes are relative to the H$\alpha$ line flux. The error of measured flux was of order $\sim$10\% for the strongest lines and $\sim$30\% for the weakest lines.

\begin{table*}
\centering
\caption{Relative emission line fluxes (Flx) of different emission regions around RX~Pup (see Fig.~\ref{fig:slits}), their radial velocities (RV), [S~II]~6716/6731 ratios, and mean radial velocities of all emission lines not including the the [O~III] lines ($<$RV$>$).  The fluxes are relative to the flux of the H$\alpha$ emission line. The radial velocities are in km/s.} \label{tab:fluxes}
\begin{tabular}{ccccccccccccccccc}
\tablewidth{0pt}
\hline
\hline
Region	&	\multicolumn{2}{c}{[O~III] 4959}	&	\multicolumn{2}{c}{[O~III] 5007}	&	\multicolumn{2}{c}{[N~II] 6548}	&	\multicolumn{2}{c}{H$\alpha$}	&	\multicolumn{2}{c}{[N~II] 6583}	&	\multicolumn{2}{c}{[S~II] 6716}	&	\multicolumn{2}{c}{[S~II] 6731}	& [S~II] & $<$RV$>$\\
& Flx & RV & Flx & RV & Flx & RV & Flx & RV & Flx & RV & Flx & RV & Flx & RV & 6716/6731\\
\hline
\decimals
A1	&	23	&	85	&	84	&	65	&	53	&	48	&	100	&	50	&	145	&	51	&	40	&	59	&	34	&	50	&	1.18	&	52$\pm$2	\\
B1	&	99	&	37	&	330	&	31	&		&		&	100	&	0	&	24	&	24	&		&		&		&		&		&		\\
B2	&		&		&	47	&	-5	&	14	&	2	&	100	&	-1	&	34	&	-15	&	16	&	-3	&	10	&	-8	&	1.56	&	-5$\pm$3	\\
C1	&	129	&	41	&	357	&	33	&		&		&	100	&	34	&		&		&		&		&		&		&		&		\\
C2	&	159	&	35	&	510	&	24	&		&		&	100	&	25	&	35	&	16	&	17	&	32	&	9.4	&	20	&	1.84	&	23$\pm$3	\\
C3	&	20	&	4	&	73	&	19	&	17:	&	-50:	&	100	&	-3	&	40	&	-21	&	13	&	-1	&	7.7	&	-16	&	1.74	&	-10$\pm$5	\\
C4	&	239	&	37	&	822	&	25	&		&		&	100	&	60	&	78	&	-22	&		&		&		&		&		&		\\
D1	&	33	&	43	&	93	&	25	&		&		&	100	&	23	&	9.7:	&	-42:	&	5.2:	&	-44:	&	4.8:	&	-34:	&	1.08:	&		\\
D2	&	83	&	15	&	255	&	20	&	15	&	18	&	100	&	17	&	41	&	14	&	12	&	-8	&	12	&	-4	&	1	&		\\
D3	&	18	&	44	&	45	&	45	&	9.6	&	-15	&	100	&	-17	&	20	&	-34	&	9	&	-14	&	7.2	&	-6	&	1.24	&	-17$\pm$5	\\
D4	&	150	&	12	&	477	&	18	&		&		&	100	&	11	&		&		&		&		&		&		&		&		\\

\hline
\end{tabular}
\end{table*}

\section{Results} \label{sec:results}
We observed an extended emission that was bright in [O~III] and H$\alpha$ east of RX~Pup. The emission forms an irregular arc extending from $\sim$30~arcsec south to $\sim$90~arcsec north-east of RX~Pup (Fig.~\ref{fig:colored}). This arc clearly shows clumps of brighter emission in the [O~III] band, while the emission in H$\alpha$ is more continuous (Fig.~\ref{fig:slits}). Moreover, a faint [O~III] emission region $\sim$10~arcsec in diameter is present $\sim$100~arcsec directly west of RX~Pup. This clump of emission seems to not be visible in our H$\alpha$ imaging. A more distant extended emission region bright in H$\alpha$ is also present around RX~Pup. While this emission is visible only east of RX~Pup, it seems to form a half-ring with a $\sim$4~arcmin radius centered at RX~Pup. The entirety of the image $\sim$1~arcmin west of RX Pup is dominated by an extended H$\alpha$ emission from a nearby H~II region that is not related to RX~Pup.

\begin{figure*}
\hspace*{\fill}\includegraphics[width=0.6\hsize]{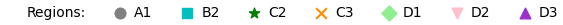}\hspace*{\fill}

\includegraphics[width=0.5\hsize]{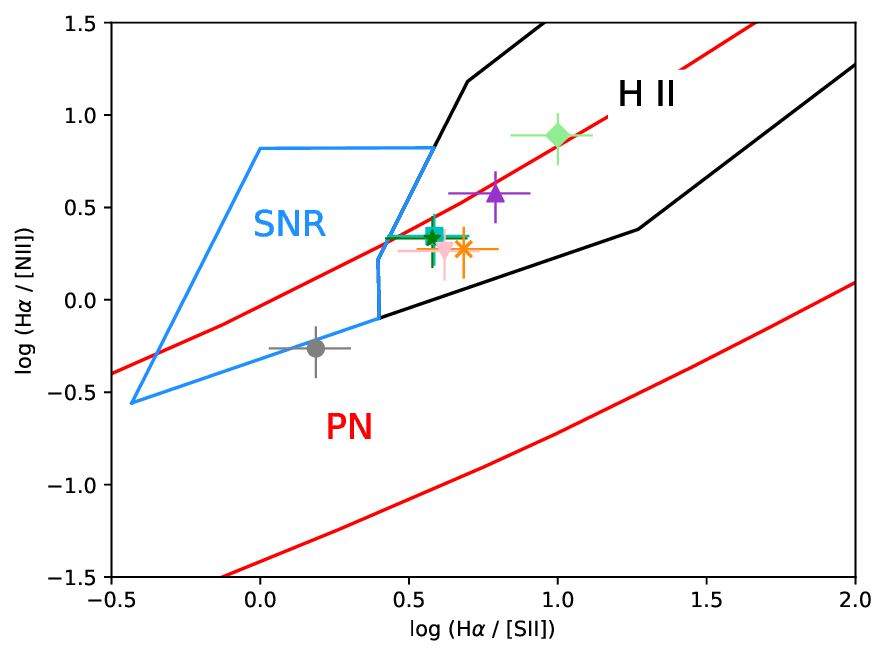}
\includegraphics[width=0.5\hsize]{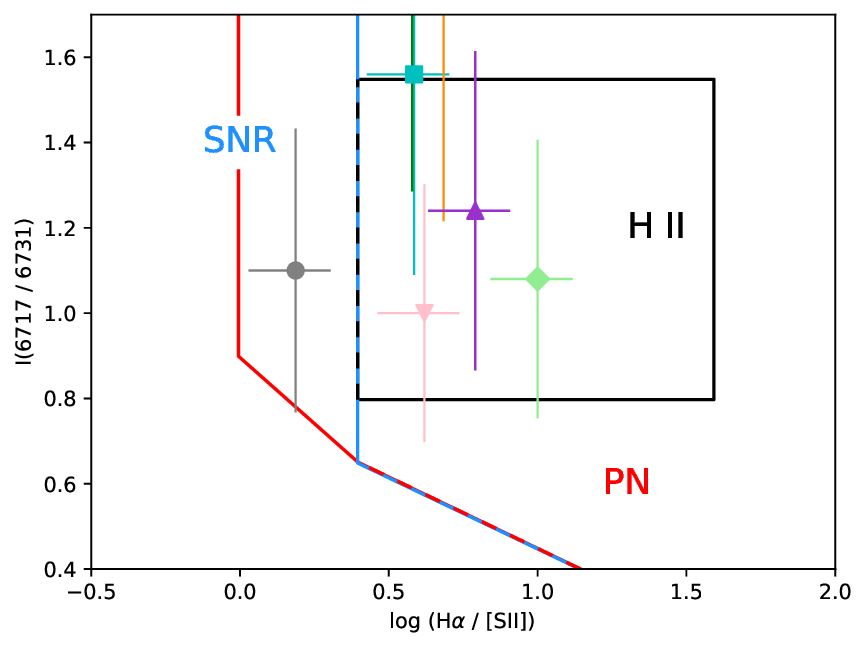}
\includegraphics[width=0.5\hsize]{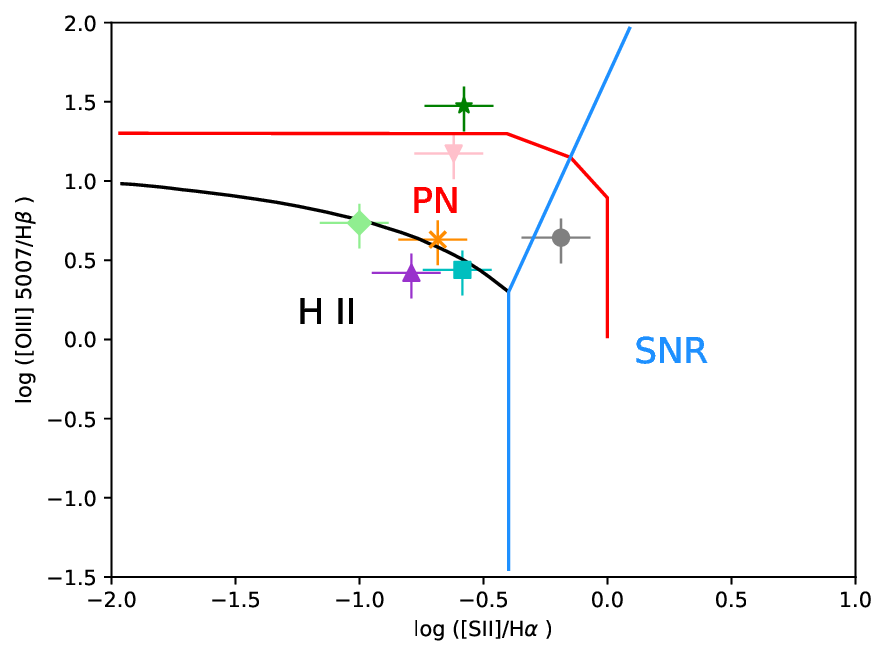}
\includegraphics[width=0.5\hsize]{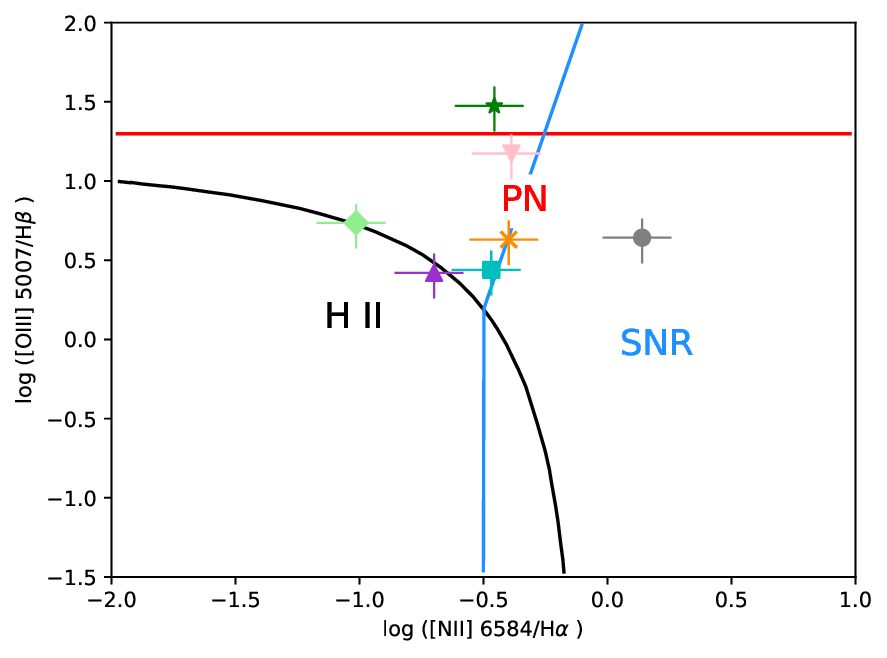}
\caption{Diagnostic diagrams based on emission line ratios designed to distinguish between H~II regions, planetary nebulae (PNe) and supernova remnants (SNR) \citep{2013MNRAS.431..279S}. The measured line ratios for the nova shell around RX~Pup are marked with colored markers. Region~A1 has emission line ratios consistent with a supernova remnant, suggesting a shock between the nova ejecta and the interstellar medium. The other regions emission line ratios are consistent with low density PNe, as expected for an old nova remnant. }
\label{fig:diagrams}
\end{figure*}

The extended emission around RX~Pup is reminiscent of a nova remnant (Fig.~\ref{fig:colored}). However, the presence of a H~II region nearby RX~Pup puts the immediate identification as a nova remnant in doubt. Diagnostic diagrams based on emission line ratios can identify different physical conditions in the emission region, which can separate out different astrophysical objects \citep[e.g.][]{2013MNRAS.431..279S}. Hence, to distinguish whether the extended emission is a nova remnant or it is associated with the H~II regions we used such diagnostic diagrams. For this we calculated the H$\beta$ intensity assuming Case~B and RX~Pup reddening E($B-V$)$\simeq$0.8~mag \citep{1999MNRAS.305..190M}. Moreover, we calculated [N~II]~6548 intensity assuming 6583/6548 line ratio of three. While nova shells are not included in the diagnostic diagrams, a region directly south of RX~Pup (region~A1) falls within supernova remnant classification region (Figure~\ref{fig:diagrams}). The supernova classification on these diagrams is based on line ratios observed in shocks, which could not be produced by a H~II region. On the other hand, it is consistent with interaction between the nova ejecta and interstellar medium, producing a shock similar to the one observed in the case of supernova remnants. Hence, we conclude that the nova remnant classification is correct. Since similar shocks produce X-ray emission we searched for X-ray emission associated with the nova ejecta of RX~Pup. In fact an X-ray source 4XMM~J081413.4-414256 is present nearby the location of the emission arc [O~III] maximum flux, at positional angle 155~degrees relative to RX~Pup \citep{2020A&A...641A.136W}. While we did not identify any optical counterparts of this X-ray source other than the irregular emission arc, a more detailed study is needed to confirm an association between 4XMM~J081413.4-414256 and the nova shell. 
The other emission regions do not show evidence of shocks. However, they overlap  the region occupied by low density planetary nebulae, as expected for an old nova remnant. We suppose that the D4 emission clump west of RX~Pup is part of the nova remnant since it is the only [O~III] emitting region within the nearby H~II region.

As for the $\sim$4~arcmin half-ring its possible identification as a nova shell is based on similarities with the irregular arc that is bright in [O~III]. More specifically, the morphology, orientation and thickness of the emission is nearly identical to the closer irregular arc already identified as a nova shell. The lack of [O~III] emission is not surprising given the fact that [O~III] flux of nova shells decreases faster than H$\alpha$ \citep{2020A&A...641A.122T}.  Moreover, both in our H$\alpha$ imaging as well as in other H$\alpha$ surveys \citep{2005MNRAS.362..689P} no similar half-rings are present in the nearby H~II region. Hence, we conclude that the $\sim$4~arcmin half-ring may be an even older nova shell of RX~Pup. However, we note that further observations are needed for a conclusive identification.

\subsection{The nova shells properties}

The fact that the nova remnants are observed mainly east of RX~Pup is consistent with observations from the 1970's outburst, where the material ejected during the nova seemed to move in an eastern direction from RX~Pup \citep{1989ApJ...337..514H}. Moreover, it is possible that the nova shells west of RX~Pup are hidden behind the nearby H~II region. Another possibility is fast deceleration of nova shells due to collision with the H~II region west of RX~Pup. In fact, RX~Pup distance of $\sim$1.6~kpc \citep{2009AcA....59..169G} is similar to a possible distance of the H~II region of 1.4--1.6~kpc \citep{2019MNRAS.483.4277C}, making this a possibility. On the other hand the emission arc is not a bowshock created due to the proper motion of RX~Pup, as its proper motion is towards the north-west direction \citep{2021A&A...649A...1G}. Asymmetric multishell structures are observed in several other symbiotic post-novae and SySt with Mira donors \citep[e.g.][]{2003ASPC..303..393C,2018A&A...616L...3B}, however in the case of RX~Pup the interpretation is more demanding due to the nearby H~II region.

\begin{figure*}
\includegraphics[width=0.5\hsize]{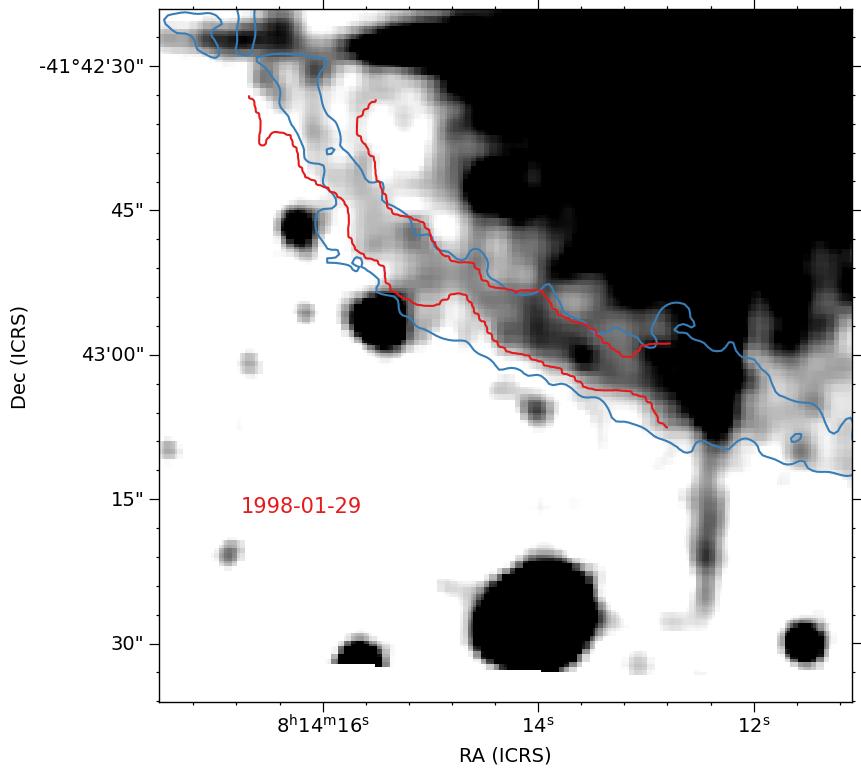}
\includegraphics[width=0.5\hsize]{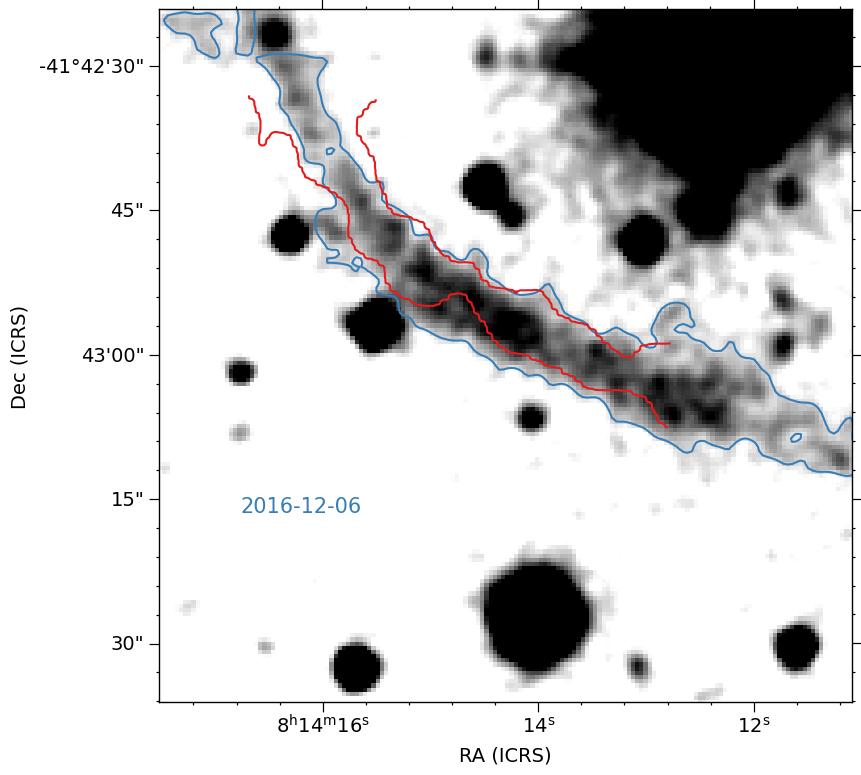}
\caption{SuperCOSMOS (left) and Swope (right) images of the irregular arc east of RX~Pup revealing expansion of the nova shell. The red contour shows position of the nova shell in the SuperCOSMOS image (1998) and the blue contour shows its position on the Swope image (2016). The SuperCOSMOS image has been smoothed with a four pixel wide Gaussian filter.}
\label{fig:expansion}
\end{figure*}

In order to date the nova shells around RX~Pup we employed archival observations from the SuperCOSMOS H$\alpha$ survey \citep{2005MNRAS.362..689P}. This image of RX~Pup was carried out on 29th of January 1998, 18.85~years before our Swope images. We revised the astrometric solution of the SuperCOSMOS image in the same way that we obtained the astrometric solution for our Swope image. On the SuperCOSMOS image the irregular arc east of RX~Pup is clearly visible, while we did not detect the $\sim$4~arcmin half-ring due to its lower relative brightness.  Visual inspection of the two images showed expansion of the nova shell (Fig.~\ref{fig:expansion}).  The distance of the irregular arc from RX~Pup on both the SuperCOSMOS and Swope images was measured in ten different directions, ranging in positional angles from 45 to 90~degrees. The positional angles were chosen to avoid blending with background stars and diffraction spikes on the SuperCOSMOS image. The distances between RX~Pup and the nova shell were measured by extracting the flux along a line at the direction of the chosen positional angle, fitting a Gauss function to the emission from the nova shell, and extracting the position of the Gauss function maximum. As a result we measured that the nova shell expanded on average by 0.65$\pm$0.16~arcsec in 18.85~years, and the average expansion rate 34.5$\pm$8.5~mas/yr. At the distance of RX~Pup this corresponds to an expansion velocity of 262$\pm$65~km/s. This is consistent with the lower limit of the 1970's nova shell expansion velocity of 80~km/s \citep{2000A&A...363..671C} and nova shell expansion velocities in other objects \citep{2020ApJ...892...60S}. 

The average radial velocities of emission lines from the irregular arc (Tab.~\ref{tab:fluxes}) are consistent with the radial velocity of RX~Pup ($\sim$20~km/s) as well as with the Galactic rotation velocity at the position and distance of RX~Pup \citep{1999MNRAS.305..190M,2000A&A...363..671C}. The radial velocities vary between regions similarly to what is observed in other SySt with outflows \citep{2008A&A...485..117S}. The radial velocities observed in regions B2, C3, and D3 (-5 to -17~km/s) are between the radial velocities of the irregular arc and the material ejected by the 1970's nova in the same, eastern direction (-80~km/s; \citealt{2000A&A...363..671C}). While they are too far away from the system (14--28~arcsec) to be associated with the 1970's outburst they may originate from yet another outburst.

The irregular shapes of nova shells around RX~Pup hinder determination of its age. Hence, for consistency we use their sizes at a positional angle of 94~degrees, where both the smaller irregular arc and the larger $\sim$4~arcmin half-ring have similar morphology and width. This gives us their sizes of 43.30$\pm$0.10 and 239.94$\pm$0.55~arcsec, respectively. Assuming constant expansion velocity of 34.5$\pm$8.5~mas/yr this corresponds to ages of $\sim$1300 and $\sim$7000~years, if the $\sim$4~arcmin half-ring is assumed to be a nova shell. We note that the material ejected in an eastern direction from RX~Pup during its last outburst expanded by 0.39~arsec in $\sim$10~years \citep{1989ApJ...337..514H}. The resulting average expansion rate of $\sim$39~mas/yr is practically the same as we have measured for the inner smaller irregular arc which indicates that the shell slowed down shortly after ejection as expected for the swept-up wind material. Moreover, the compatibility of both estimates implies that the shells remain in the free expansion phase for at least $\sim$1000 years. This is consistent with the observed CN shells which show constant expansion velocities for at least the first century after the outburst \citep{2020ApJ...892...60S,2024A&A...681A.106C} even when strong interaction between the nova ejecta and interstellar medium is present \citep{2012AJ....143..143S,2012ApJ...761...34L,2016A&A...595A..64H}.

\section{Conclusions} \label{sec:concl}

In this work we reported the discovery of a $\sim$1300 year old nova shell forming an irregular arc around RX~Pup. Additionally, we identified a possible $\sim$7000~year nova shell appearing as a $\sim$4~arcmin half-ring. Together with the 1970's outburst this implies a change of nova recurrence time from $\sim$5700 to $\sim$1250~years assuming constant expansion rate. A long timescale and low luminosity of the last outburst of RX~Pup suggests a  WD mass in the range of 0.6--0.8~M$_\odot$ \citep{1999MNRAS.305..190M,2010arXiv1011.5657M}. If we assume typical values of a mass of the WD M$_{\mathrm{WD}}=0.65$~M$_\odot$ and and a temperature of its isothermal core T$_{\mathrm{WD}}=30\times 10^{6}$~K a change of mass transfer rate can be estimated by interpolating the theoretical classical nova models of \citet{2005ApJ...623..398Y}. This approach suggests that a change in recurrence time of nova outbursts from $\sim$5700 to $\sim$1250~years for the assumed WD parameters is due to a change of the average mass transfer rate from $\sim 2.8\times10^{-8}$~M$_\odot$/yr to $\sim 7.9\times10^{-8}$~M$_\odot$/yr. Hence, the mass transfer rate in RX Pup increased by roughly a factor of three. Since the WD had to accumulate mass on its surface before the outburst that occurred $\sim$7000~years ago, this means that the measured change in mass transfer rate occurred on a timescale of roughly 10,000 years.  Moreover, regions C3 and D3 (Fig.~\ref{fig:slits}) are not visible in our imaging but their spectra suggest that they may originate from yet another outbursts. If confirmed, this would suggest that the mass transfer rate evolution in RX~Pup was faster than we estimated.

The change of mass transfer rate is the first measurement of variable mass transfer rate in binary stars on timescales of millennia. This result is consistent with the evolutionary status of the Mira in RX~Pup implied by its dust obscuration events \citep{2001ASSL..265..227M}. Namely, single stars similar to the mass-donor in RX~Pup experience variable mass loss on similar timescales \citep{2002MNRAS.334..498Z}. Hence, RX~Pup may serve as a laboratory of how the evolution of the mass donor influences the mass transfer rate in binary stars with wind accretion. This is  particularly interesting given that such binaries have been suggested as promising supernova Ia candidates \citep{2019MNRAS.485.5468I}.

\section*{Acknowledgements}
This work was supported by Polish National Science Center grant Sonatina 2021/40/C/ST9/00186. JM acknowledges support from the Polish National Science Center grant 2019/35/B/ST9/03944 and from the Copernicus Academy grant CBMK/01/24. The paper is based on spectroscopic observations made with the Southern African Large Telescope (SALT) under programmes 2019-1-SCI-025 and 2019-2-SCI-021 (PI: I{\l}kiewicz). Polish participation in SALT is funded by grant No. MEiN 2021/WK/01. This research made use of APLpy, an open-source plotting package for Python \citep{2012ascl.soft08017R,robitaille_thomas_2019_2567476}.

\bibliography{literature}{}
\bibliographystyle{aasjournal}

\appendix

\begin{table*}[h]
\centering
\caption{Log of spectroscopic observations at different slit positions (see Fig.~\ref{fig:slits}).} \label{tab:speclog}
\begin{tabular}{cccc}
\tablewidth{0pt}
\hline
\hline

Slit position & Date &  JD & Exposure time [s]\\
\hline
\decimals
A & 2019-11-02 & 2458789.5 & 2510\\
B & 2019-10-05 & 2458761.6 & 2400\\
C & 2019-11-04 & 2458791.5 & 2510\\
D & 2019-11-03 & 2458790.5 & 2510\\
\hline
\end{tabular}
\end{table*}

\begin{figure}
    \centering
    \includegraphics[width=\columnwidth]{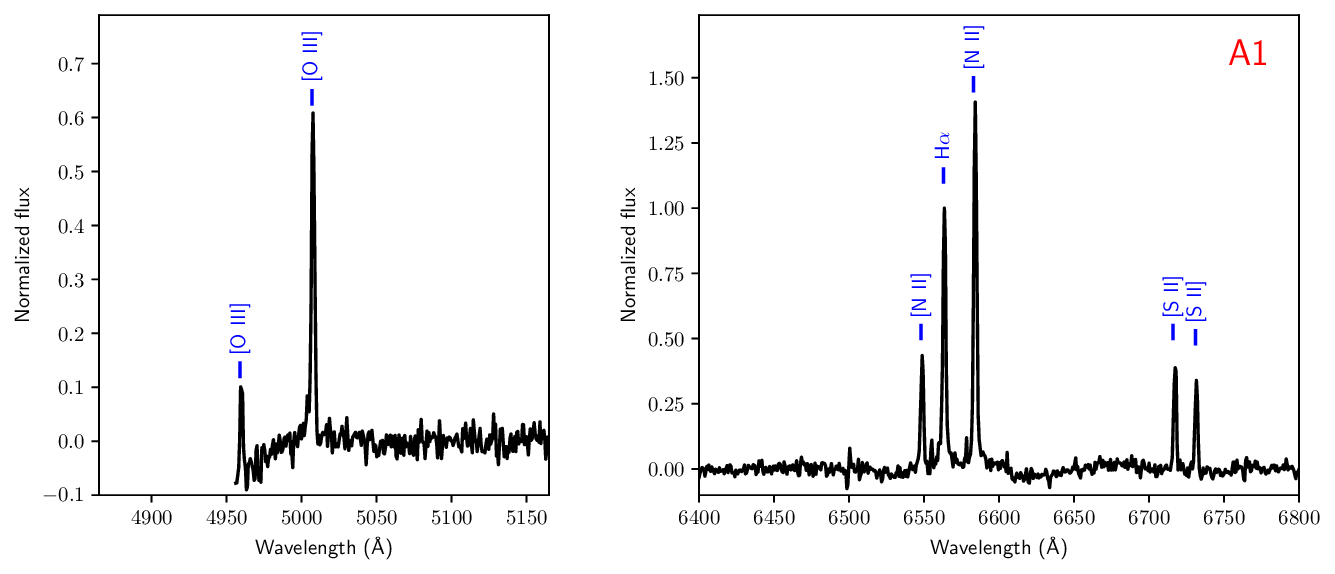}
    \includegraphics[width=\columnwidth]{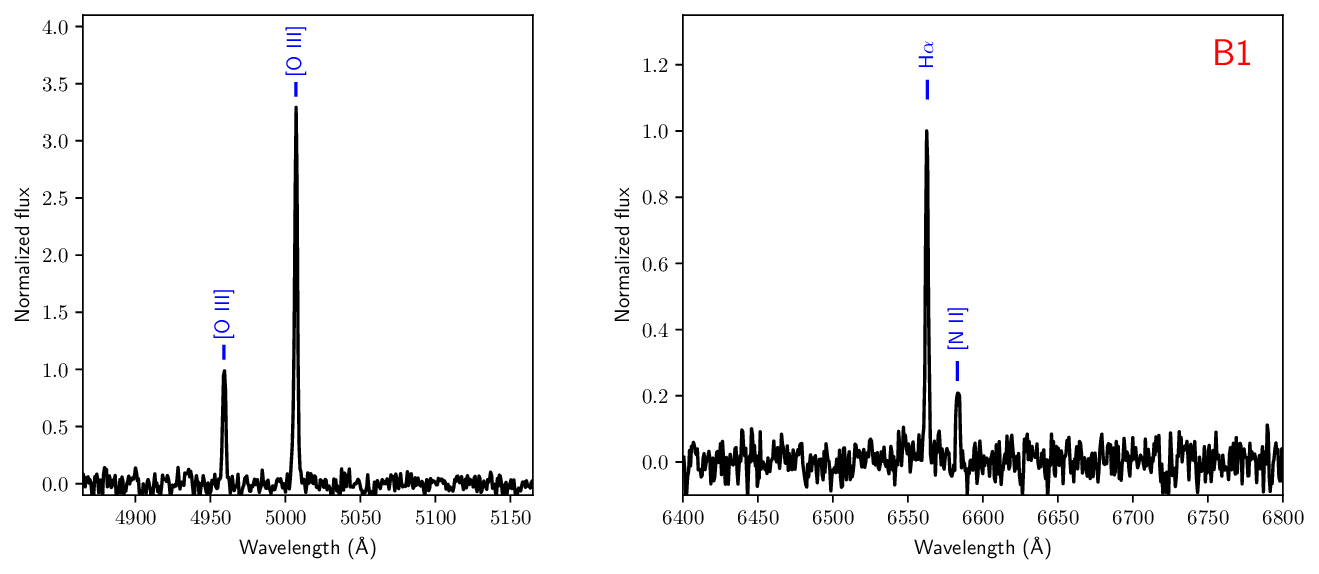}
    \includegraphics[width=\columnwidth]{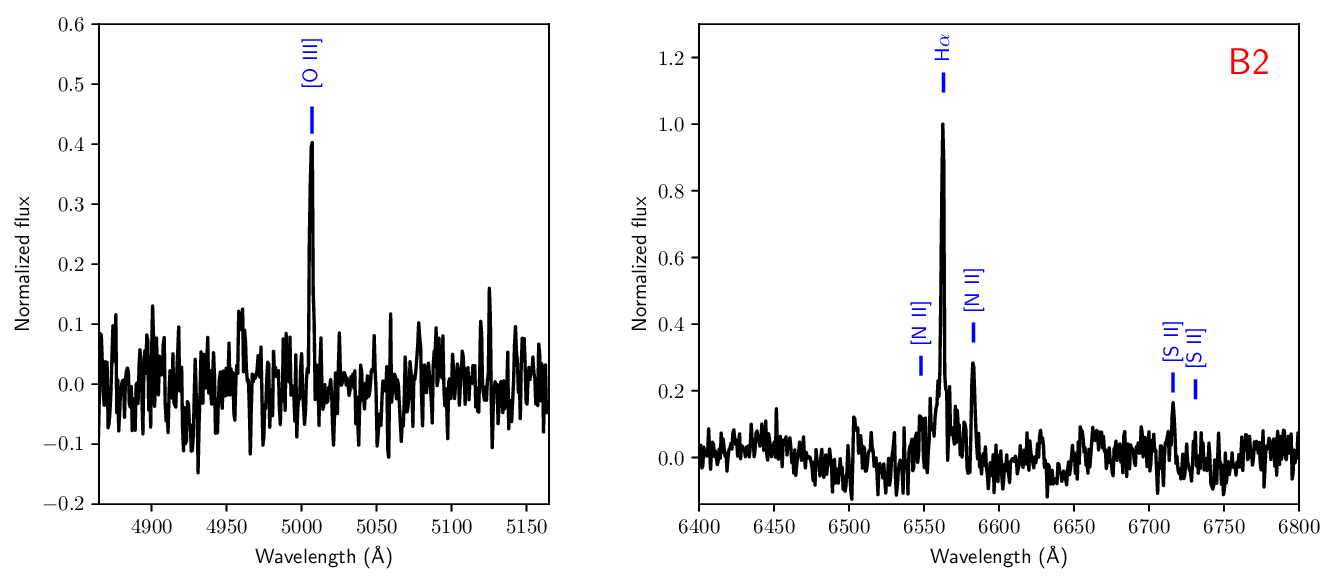}
    \caption{Spectra of different nova shell regions in the H$\alpha$ and [O~III] regions with names of regions given with a red text (see Fig.~\ref{fig:slits}). The flux was normalized to the H$\alpha$ maximum flux.}
    \label{fig:spectra}
\end{figure}

\begin{Contfigure}
    \centering

    \includegraphics[width=\columnwidth]{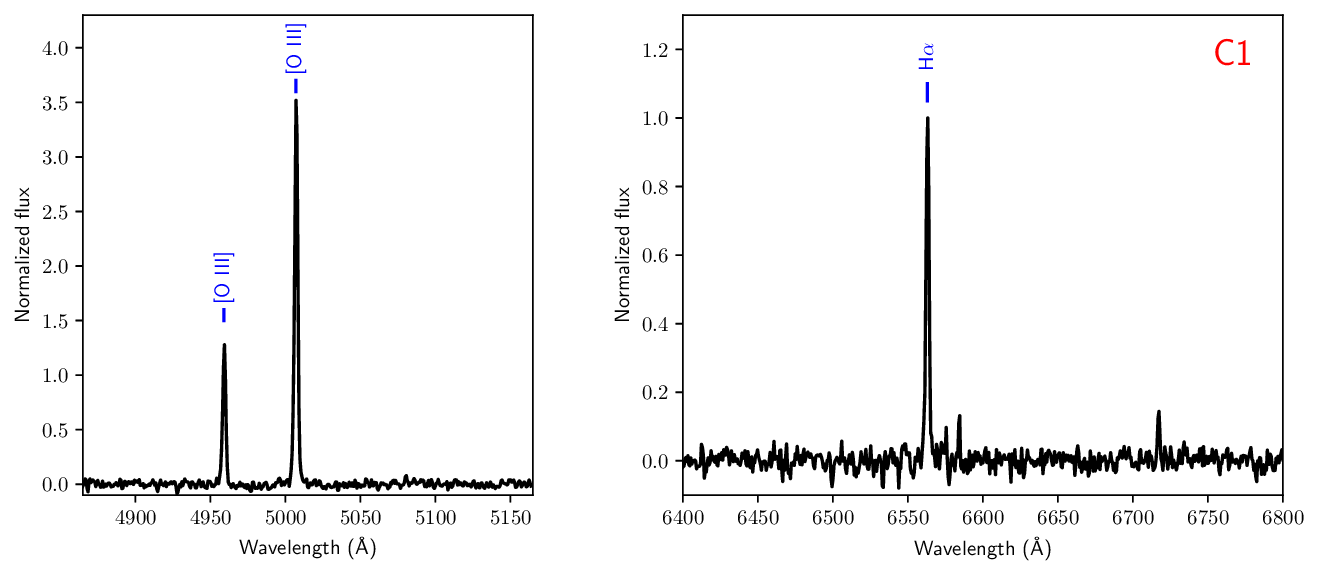}
    \includegraphics[width=\columnwidth]{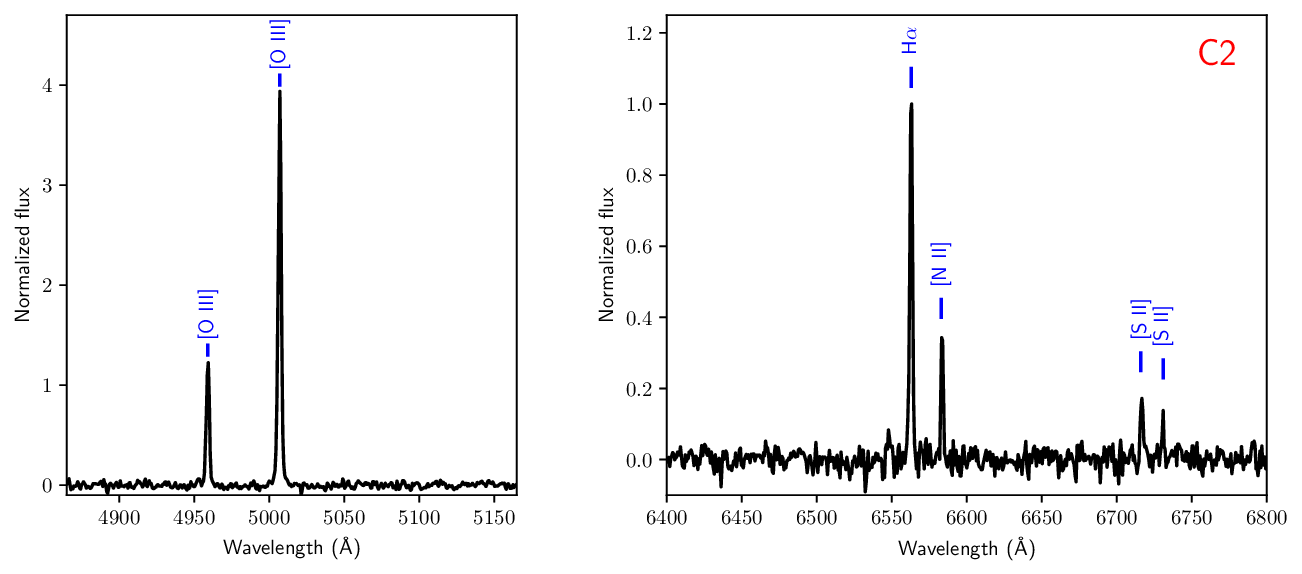}
    \includegraphics[width=\columnwidth]{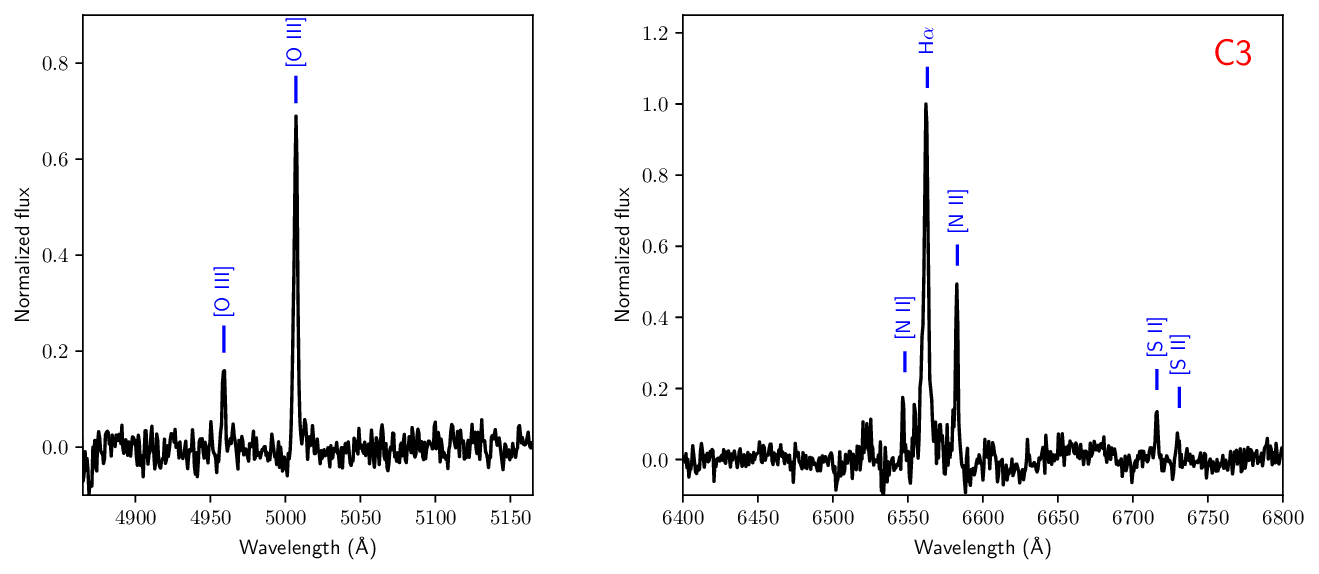}
    \caption{Continued.}
\end{Contfigure}

\begin{Contfigure}
    \centering

    \includegraphics[width=\columnwidth]{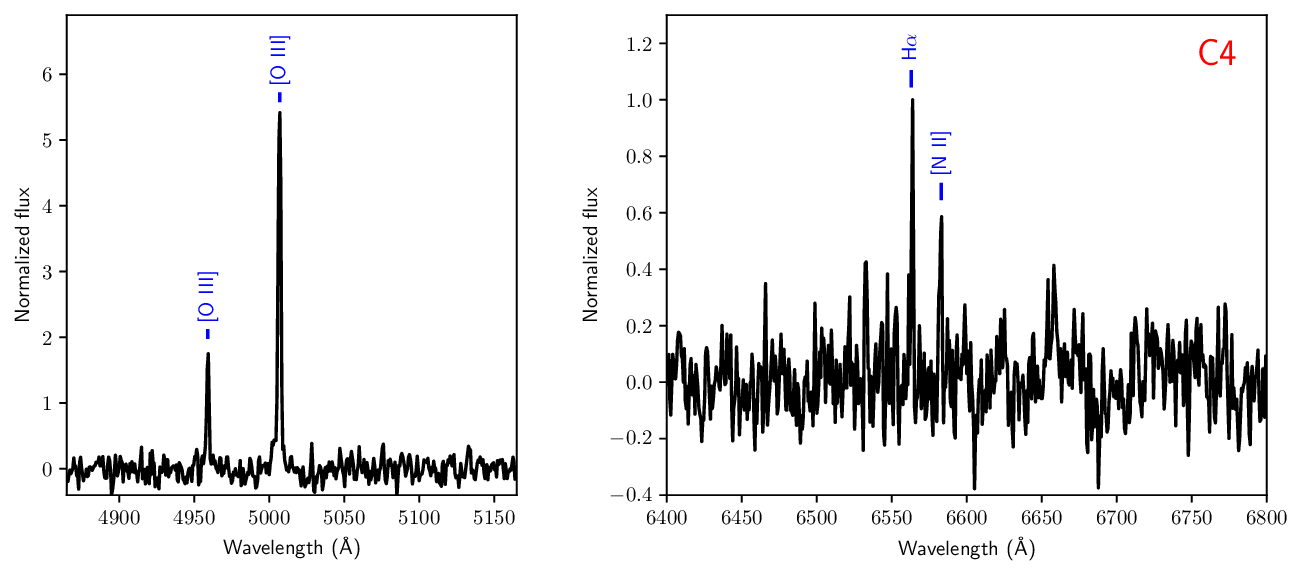}
    \includegraphics[width=\columnwidth]{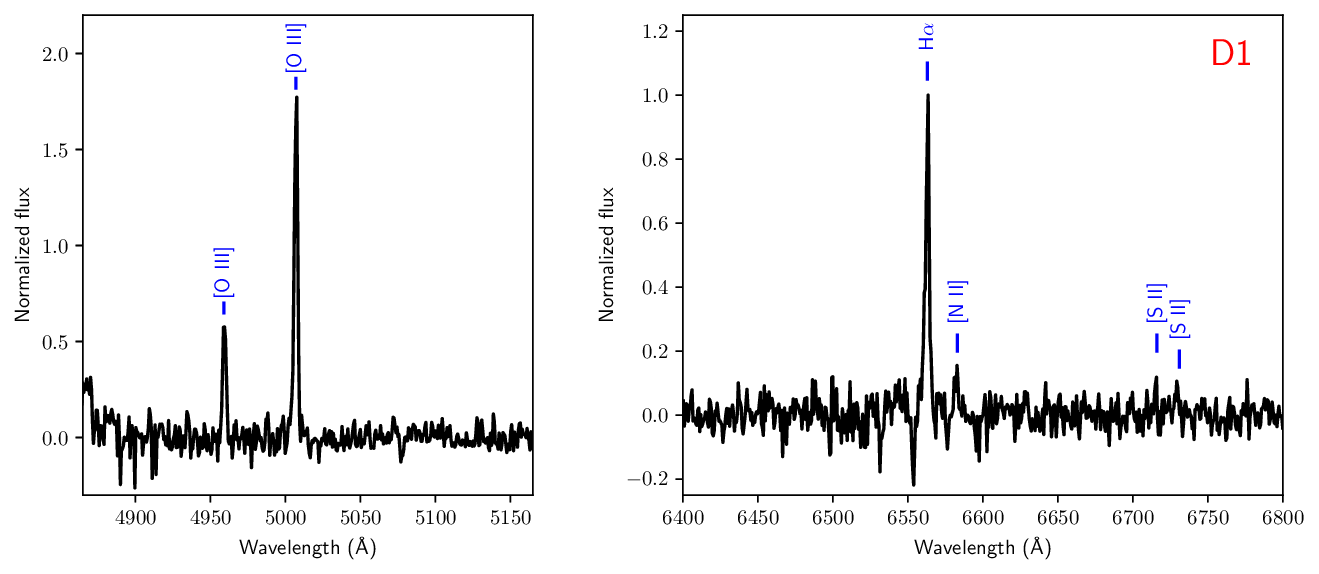}
    \includegraphics[width=\columnwidth]{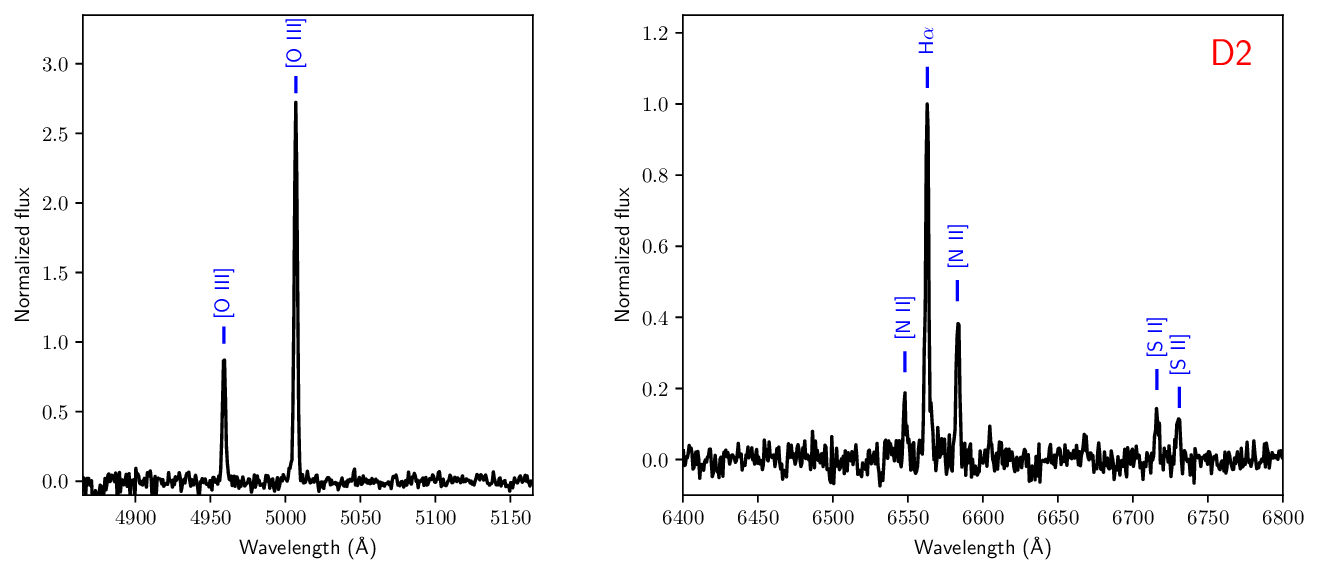}
    \caption{Continued.}
\end{Contfigure}

\begin{Contfigure}
    \centering

    \includegraphics[width=\columnwidth]{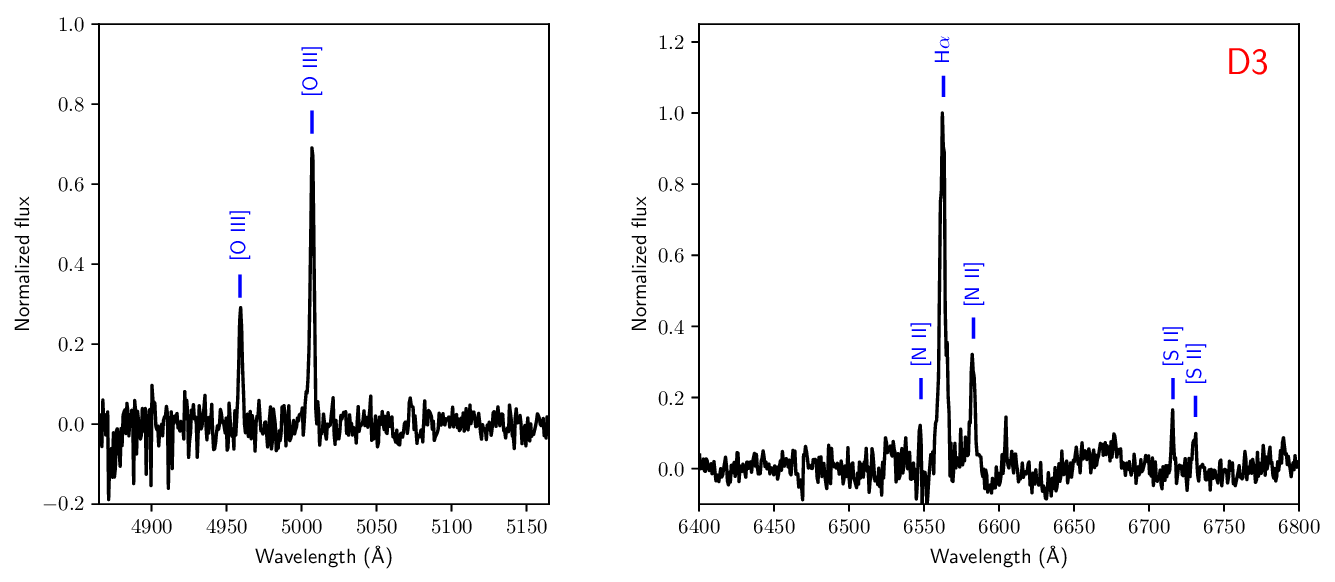}
    \includegraphics[width=\columnwidth]{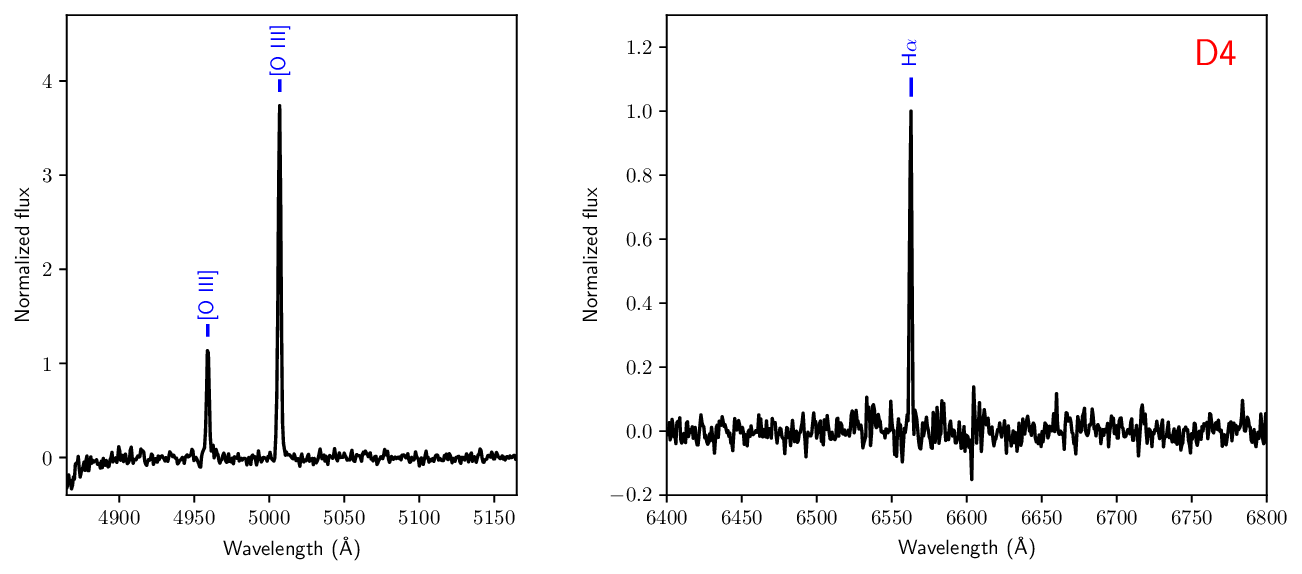}
    \caption{Continued.}
\end{Contfigure}

\end{document}